\documentclass[preprint,1p, times,authoryear]{elsarticle}

\usepackage{amssymb}

\usepackage{tabularx}

\newenvironment{conditions*}
  {\par\vspace{\abovedisplayskip}\noindent
  \tabularx{\columnwidth}{>{$}l<{$} @{${}={}$} >{\raggedright\arraybackslash}X}}
  {\endtabularx\par\vspace{\belowdisplayskip}}

\usepackage{enumerate}
\usepackage{nicefrac} 
\usepackage{enumitem}
\usepackage{caption} 
\usepackage{multirow}
\usepackage{booktabs}
\usepackage{amsfonts}
\usepackage{amsmath}

\journal{Journal of Empirical Finance}

\usepackage{hyperref}
\usepackage{academicons}
\usepackage{xcolor}

\usepackage{orcidlink}
\begin{document}

\begin{frontmatter}


\tnotetext[label1]{Declarations of interest: none}

\title{Nowcasting Stock Implied Volatility with Twitter}

\affiliation[inst1]{organization={Department of Statistics and Risk, KU Leuven},
            addressline={Celestijnenlaan 200B},
            city={Leuven},
            postcode={3000},
            country={Belgium}}

\affiliation[inst2]{
            organization={Department of Computer Science, KU Leuven},          addressline={Celestijnenlaan 200A},
            city={Leuven},
            postcode={3000},
            country={Belgium}}
\author[inst1,inst2]{Thomas Dierckx\corref{cor1} \orcidlink{0000-0001-9585-0566} }
\ead{thdierckx@gmail.com}
\author[inst2]{Jesse Davis \orcidlink{0000-0002-3748-9263}}
\ead{jesse.davis@kuleuven.be}

\author[inst1]{Wim Schoutens \orcidlink{0000-0001-8510-1510}}
\ead{wim.schoutens@kuleuven.be}

\cortext[cor1]{Corresponding author}

\begin{abstract}
In this study, we predict next-day movements of stock end-of-day implied volatility using random forests. Through an ablation study, we examine the usefulness of different sources of predictors and expose the value of attention and sentiment features extracted from Twitter. We study the approach on a stock universe comprised of the 165 most liquid US stocks diversified across the 11 traditional market sectors using a sizeable out-of-sample period spanning over six years. In doing so, we uncover that stocks in certain sectors, such as Consumer Discretionary, Technology, Real Estate, and Utilities are easier to predict than others. Further analysis shows that possible reasons for these discrepancies might be caused by either excess social media attention or low option liquidity. Lastly, we explore how our proposed approach fares throughout time by identifying four underlying market regimes in implied volatility using hidden Markov models. We find that most added value is achieved in regimes associated with lower implied volatility, but optimal regimes vary per market sector. 
\end{abstract}

\begin{highlights}
\item Next-day movements in stock implied volatility can be predicted using random forests.
\item Attention and sentiment features extracted from Twitter improve predictive performance.
\item Predictive performance varies significantly across the 11 traditional stock market sectors.
\item Implied volatility regimes identified by hidden Markov models provide actionable insights on when the proposed approach works best per stock market sector.
\end{highlights}

\end{frontmatter}




\section{Introduction}
Today's age is characterized by an ever-increasing connected and opinionated world. The widespread adoption of social media has caused significant changes in the world across many domains, and more are probably to follow. In the case of financial markets, participants now have access to countless online platforms to share their thoughts and feelings on certain events. Proponents of the Efficient Market Hypothesis \citep{fama} ought to be pleased. The advent of mass media facilitates rapid information diffusion, possibly propelling markets into a higher tier of price efficiency. However, behavioral economists would argue that this type of media might very well influence investors and incite herd behavior which in turn induces inefficiency \citep[e.g.][]{baker2006,herd2}. Theory aside, this new wealth of information has not escaped the notice of the financial establishment. Indeed, data providers such as Bloomberg and Refinitiv now offer extensive social media indicators to help financial institutions navigate this new world. Although the competitive edge that resides in these alternative data sources remains veiled in secrecy, an abundance of academic studies have already tried quantifying the interplay between social media and certain financial variables, providing insight into the predictive power of the masses. 

Existing research has mainly focused on the Twitter platform and its influence on three prominent financial variables: stock price \citep[e.g.][]{GROKLUMANN2019171, SCHNAUBELT2020103895}, realized volatility \citep[e.g.][]{Karagozoglu}, trading volume \citep[e.g.][]{twitliq} or a combination thereof \citep[e.g.][]{Oliveira2017, twitnoise}. 
Remarkably, current literature completely overlooks the interaction between social media and the market implied volatility of stocks. Derived from option prices, this variable is deemed to be one of the more important parameters in the world of derivatives. In contrast to historical volatility, this is a forward-looking metric that indicates how much risk the market expects a certain asset to exhibit in the coming period. As this variable serves as a proxy for both market sentiment and option contract prices, the ability to predict its movements would be advantageous for the practice of asset management and market making alike.
 
Most prevailing studies in this domain have two important methodological shortcomings. First, analysis is typically performed on either a handful of arbitrarily chosen stocks or indices tracking the entire market \citep[e.g.][]{GROKLUMANN2019171}, omitting sector idiosyncrasies in the process. Second, hypotheses are commonly tested on a relatively small time window ranging from a month \citep[e.g.][]{bollen} to a few years \citep[e.g.][]{SCHNAUBELT2020103895}. However, the ever-changing nature of financial markets warrants a closer look into how the interplay between social media patterns and financial variables evolves over longer periods of time. As patterns may emerge and dissipate over time, a crucial aspect of analysis is often left out. 

The contribution of this study is threefold. First, to the best of our knowledge, we are the first to investigate to what extent a stock its one-day ahead movement in implied volatility can be predicted using machine learning on different combinations of feature sources, including Twitter. Second, instead of arbitrarily choosing a handful of stocks for our study, we diversified our stock universe across the 11 traditional US stock market sectors yielding 165 stocks in total. This allowed us to measure and explore the variability in predictive performance present among sectors. Lastly, we examined predictive performance on an out-of-sample period spanning January 1st, 2013 till March 1st, 2019. This period is significantly larger than many other studies and gave us the opportunity to not only be more robust, but also examine predictive performance throughout time. Instead of performing a year by year analysis of predictive performance, we used hidden Markov models to identify four regimes in the implied volatility of a stock and gauged whether performance varies across them. We argue that this alternative quantification of time yields more actionable insight as it allows practitioners to better anticipate future performance.

\section{Preliminaries}
This section presents background information on the key components used in this study. First, Section \ref{implied_section} explains market implied volatility and its relation to the world of derivatives. Second, Section \ref{random_forest} describes the random forest machine learning model which is used to perform predictions. Lastly, Section \ref{hmm} describes the hidden Markov model which is used to quantify regimes in market implied volatility. 

\subsection{Market Implied Volatility} 
\label{implied_section}
In the world of derivatives, options are one of the most prominent types of financial instruments. As sellers of options are exposed to risk for the duration of the contract, they want to be properly compensated. Measuring this risk requires considering the expected price fluctuations of the underlying asset over the duration of the contract. 
This expectation is better known as implied volatility and it varies with the strike price and duration of an option contract. To obtain a more general measure, the implied volatility of option contracts that expire on the same date can be combined into a single implied volatility measure. A famous example of this is the CBOE Volatility Index, which combines the implied volatility of different option contracts on the SPX into an index that is better known as the VIX. 

More concretely, the VIX is a measure of expected price fluctuations in the S\&P 500 Index over the next 30 days. It is famously known as the fear index and is considered a reflection of investor sentiment on the condition of the market. Equation \ref{eq:vix}, taken from the VIX white paper \citep{VIX}, shows how to compute the VIX for a given term $T$:

\begin{equation}
  \mathit{VIX} = 100 \times \sqrt{\frac{2}{T} \sum\limits_{i} \frac{\Delta K_{i}}{K_{i}^{2}} e^{RT} Q(K_{i}) - \frac{1}{T} \bigg[ \frac{F}{K_{0}} - 1 \bigg]^{2}}
  \label{eq:vix}
\end{equation}
\noindent where:
\begin{conditions*}
 T    &  is time to expiration \\
 F    &  is the forward index level derived from the index option prices \\
K_{0} &  is the first strike below the forward index level F \\
K_{i} &  is the first strike price of the $i^{th}$ out-of-the-money option: a call if $K_{i}>K_{0}$, a put if $K_{i}<K_{0}$, or both put and call if $K_{i}=K_{0}$ \\
R & is the risk-free rate to expiration \\
\Delta K_{i} & is the interval between strike prices \\
Q(K_{i}) & is the midpoint of the bid-ask spread for each option with strike $K_{i}$ \\
\end{conditions*}

The equation for computing the VIX is applicable to any asset where option contracts are available. Although this measure can be calculated for any arbitrary term, the duration of the option contracts will seldom match the chosen term $T$. Indeed, option contracts typically have fixed expiration dates and there is no guarantee that there are option contracts available with a duration equal to the given term. To overcome this obstacle, the VIX is first calculated for the option contracts expiring right before and after the desired target date. The VIX for the given term can then be calculated by linearly interpolating between the two computed measures, as outlined in \citep{VIX}.

\subsection{Random Forests} 
\label{random_forest}
Random forests \citep{rf} are a popular machine learning approach for learning a predictive model. They consist of multiple different decision (or regression) trees whose predictions are combined into one final prediction. The combination is typically done by taking the mode (or average) of all outputs. While several variations exist for learning a random forest, all of them are relatively straightforward. We summarize one of many popular procedures. Given data $D = \{(x_{i}, y_{i})\}_{i=1}^{n}$, where each $x_{i}$ has $F$ features and for $k= 1 \dots K$ trees: 

\begin{enumerate}[noitemsep]
    \item Obtain subset $d$ by sampling $m<n$ examples with replacement from $D$.
    \item Train a decision tree on $d$ using a random subset features $f \subseteq F$, i.e. using CART \citep{cart}.
\end{enumerate}
The prediction for a regression problem can then be obtained by:
\begin{equation}
     \hat{y}_{i} = \frac{1}{K} \sum_{k=1}^{K} f_{k}(x_{i})
    \label{rf1}
\end{equation}
where $f$ is a function in the set of all possible decision trees and $K$ is the total number of trees in the ensemble.

The advantages of random forests include that they are fast to build, are not affected by feature scaling,  are robust to irrelevant predictors and noisy data \citep{noisebag}. Moreover, their method of constructing an ensemble model by randomly subsampling both data points and features during the learning process helps decorrelate the predictions made by the individual trees, which in turn reduces overfitting on the training data.





\subsection{Hidden Markov Models} 
\label{hmm}

Hidden Markov models (HMM) are a generative approach for modeling systems that follow a Markov process \citep{hmm}.
The main assumption is that while this process $Z$ is hidden, it can be learned from an observable sequence $X$ whose behaviour depends on $Z$. More formally, the HMM models the joint distribution of a sequence of hidden states $Z$ and observations $X$ described by:

\begin{equation}
    P(Z_{1:K}, X_{1:K}) = P(Z_{1})P(X_{1}|Z_{1}) \prod_{t=2}^{K} P(Z_{t}|Z_{t-1})P(X_{t}|Z_{t})
    \label{eqhmm1}
\end{equation}

Given the number of hidden states $K$ and observed sequence $X$, the model is fully determined by its parameters $\pi$, $A$, and $B$ which represent the initial state distribution, state transition model, and emission probabilities model, respectively. The initial state distribution is a $K \times 1$ vector denoting the probabilities that the process is each of the $K$ states in the first timestep. The transition model is a $K \times K$ stochastic matrix where each element $A_{i,j}$ denotes the probability of transitioning from state $Z_{t-1, i}$ to $Z_{t, j}$ where $i, j \in \{1, \dots, K\}$. Lastly, the emissions probability model is a $M \times K$ matrix, with $M$ representing the number of distinct observations, whose elements $B_{k, j}$ denote the probability of observing $X_{t, k}$ given state $Z_{t, j}$.

The three key tasks associated with hidden Markov models are: 
\begin{enumerate}[noitemsep]
    \item What is the probability that a sequence of observations $X$ was generated by a given HMM?
    \item Given an HMM, what sequence of hidden states $Z$ best explains a given sequence of observations $X$?
    \item Given a sequence of observations $X$, learn an HMM with parameters $\pi$, $A$, and $B$ that would generate them.
\end{enumerate}

The first two tasks can be solved using dynamic programming using the forward-backward algorithm~\citep{fwdback} and Viterbi \citep{viterbi} algorithm, respectively. The third problem is solved by the Baum-Welch algorithm \citep{baumwelch} which uses an iterative expectation-maximization approach.

\section{Methodology}

The main goal of this study is to explore the following questions:
\begin{enumerate}[noitemsep]
    \item To what extent are one-day ahead movements in end-of-day implied volatility predictable, and do features extracted from Twitter improve performance?
    \item Does performance vary across the 11 different stock market sectors? If so, are there any obvious factors that might explain this variability?
    \item Can we identify underlying market regimes in implied volatility that influence the performance of our proposed approach?
\end{enumerate}

We tackle the first question by performing an ablation study using random forests on feature configurations including stock price, stock implied volatility, and Twitter features. The study encompasses a universe of 165 stocks over an out-of-sample period spanning January 1st, 2013 till March 1st, 2019. To examine the second question, we diversified our stock universe over the 11 traditional stock sectors and grouped predictive performance by stocks belonging to the same sector. The third and last question was studied by using a hidden Markov model to identify four distinct implied volatility regimes per stock, after which predictive performance was grouped by regime. 

The next few sections explain our methodology in more detail. First, Section \ref{universe} outlines the stock universe we used for our study. Second, Section \ref{meth:feats} explains how we obtained the relevant data for each stock and how we constructed features for prediction. Section \ref{meth:pred} and Section \ref{meth:eval} then respectively show how we used machine learning to predict our target variable and how we evaluated the performance of the approach. Lastly, Section \ref{meth:hmm} explains how we used hidden Markov models to identify regimes in implied volatility which we later use to evaluate our prediction performance through time.

\subsection{Stock Universe Selection}
\label{universe}

In order to obtain a diversified universe of stocks, we looked at the popular SPDR and Vanguard Electronic Traded Funds (ETF) that track the 11 traditional US stock market sectors. For each sector, we selected the 15 most liquid stocks based on their average daily dollar-weighted option volume for a total of 165 stocks. Some stocks were excluded due to stock splits (i.e. we kept GOOG and dropped GOOGL), a late introduction to the stock market (i.e. PYPL, ROKU, and SNAP only got introduced after 2015), and ambiguous names making it hard to obtain relevant tweets (i.e. DOW is a chemical company but also a common alias for the Dow Jones Index). Note that we replaced the excluded stocks to maintain 15 stocks per sector for our study. Table \ref{tab:symb} provides a concise overview of our stock universe. Refer to \ref{appA} for a full overview of which stocks were selected per market sector. 

\begin{table}[!ht]
\setlength{\tabcolsep}{10pt} 
\caption{This table presents the 11 different stock market sectors together with their corresponding SPDR ETF symbol and number of stocks considered in this study. The symbols are used to denote sectors throughout this paper, but are not indicative of stocks only belonging to the SPDR ETF portfolio. }
  \centering
\begin{tabular}{ccc}
\hline
\textbf{Symbol} & \textbf{Sector}        & \textbf{Selected Stocks} \\
\hline
XLC             & Communication Services & 15 \\
XLY             & Consumer Discretionary & 15 \\
XLP             & Consumer Staples       & 15 \\
XLE             & Energy                 & 15 \\
XLF             & Financials             & 15 \\
XLV             & Healthcare             & 15 \\
XLB             & Materials              & 15 \\
XLI             & Industrials            & 15 \\
XLK             & Technology             & 15 \\
XLRE            & Real Estate            & 15 \\
XLU             & Utilities              & 15 \\
\hline
\end{tabular}
\label{tab:symb}
\end{table}

\subsection{Data Acquisition and Feature Generation}
\label{meth:feats}

We consider data ranging from January 1st, 2011 through March 1st, 2019 for three data sources:
\begin{enumerate}[noitemsep]
\item \textbf{Stock price data}  which consists of historical end-of-day adjusted closing prices for each stock in our universe downloaded from Yahoo Finance.
\item \textbf{Option contract price data} which consists of historical end-of-day option chains for each stock in our universe obtained from IVolatility.
\item \textbf{Twitter data} which consists of all relevant tweets published for each stock. These were collected by filtering on \textit{cashtags}, which are popular string identifiers authors use to indicate their message is about a certain stock (i.e. a tweet about the Apple stock typically contains \$AAPL). In contrast to other research, we did not employ additional filtering techniques to discard potential spam. Most additional filtering rules appear arbitrary and there seems to be no clear evidence of their validity. 
\end{enumerate}

In total, four features were extracted per stock for each trading day. First, we simply used the end-of-day adjusted closing price from the stock price data. Second, we calculated the end-of-day 30-day implied volatility using the VIX method on the option contract data. Third and last, we derived two numerical features from our textual Twitter corpus: end-of-day total tweet publication count and end-of-day average sentiment polarity. The former represents the total number of published tweets on a given day. The latter was obtained by performing sentiment analysis using VADER \citep{vader}, a lexicon- and rule-based sentiment model that is specifically well-tailored to social media text, on individual tweets. This yields a sentiment polarity score $s \in [-1, 1]$ for each tweet, which was then used to compute the daily average. 

In an effort to capture temporal information residing in the original feature timeseries, we generated two additional predictors per feature. To this end, the daily difference (or first-order difference) and the difference between the daily value and its exponential moving average of the last 10 trading days was taken. Table \ref{tab:features} outlines the different data sources and their features used in this study. Note that the original adjusted closing price was omitted, as this is typically a non-stationary variable offering little value to a prediction model.

\begin{table}[!ht]
\setlength{\tabcolsep}{6pt}
 \caption{This table provides a summary of the features considered per data source. The first row indicates what original features were extracted, whereas the last three rows indicate (*) which features were considered for the actual study.  Note that the last two rows denote a specific feature engineering technique applied to the original feature. }
  \centering
  \begin{tabular}{cccc}
    \toprule
    & Stocks  & Options & Twitter \\
    \midrule
    Extracted & Adj. Closing Price & Implied Volatility & Count, Sentiment \\
    \midrule
    Original &      & * & * \\
    1st Order Diff. & * & * & * \\
    EMA(10) Diff. & * & * & * \\

    \bottomrule
  \end{tabular}
  \label{tab:features}
\end{table}

\subsection{Predicting Movements in Implied Volatility}
\label{meth:pred}
This study aims to predict one-day ahead movements in a stock's 30-day implied volatility. Concretely,  given information at the end of trading day $t$, we predict whether implied volatility will have moved up or down by the end of next trading day $t+1$. To do so, we construct a binary target variable for day $t$ as:

\begin{equation}
  y_{t}=
  \begin{cases}
    1, & \text{if $(\mathit{ivolatility}_{t+1}-\mathit{ivolatility}_{t})>0$}.\\
    0, & \text{otherwise}.
  \end{cases}
  \label{eq:target}
\end{equation}
where $\mathit{ivolatility_{t}}$ denotes the end-of-day implied volatility on day $t$.

In order to predict our target variable, we used random forest classifiers even though more powerful models may exist. For example, the highly popular gradient boosted trees \citep{gbm} have been shown to generally perform slightly better than random forests \citep{boostbest, boostbest2}. However, they are very sensitive to hyper-parameter configurations and require longer runtimes for training. The main goal of this study is not to maximize predictive performance, but rather probe the feasibility of our proposed approach. In addition, it has been suggested that random forests might generally work better on noisy data \citep{noisebag}, which is especially convenient when working on financial data. Lastly, we did not consider techniques from the domain of deep learning due to the complexity of the models and relatively small number of data points in our study. 

Ultimately, we used 64 distinctive random forest configurations built using Sklearn. Each random forest was built with 1000 trees and a unique combination of different hyper-parameters that control maximum tree depth, the minimum number of samples required to split an internal node, and the minimum number of samples required to be in a leaf node. Each individual tree was built by sampling the training dataset (with replacement) and only considering a random number of $\sqrt{f}$ features where $f$ denotes the total amount of features. The models were trained on a temporally ordered feature matrix $X$ of dimension $T \times K$, obtained by using any subset of features $K$ from Section \ref{meth:feats} and period $T$. Table \ref{tab:rfparams} specifies the possible random forest configurations considered in this study.
\begin{table}[!ht]
\setlength{\tabcolsep}{10pt} 

\caption{This table presents the different possible values considered for different hyper-parameters available in the random forest implementation of Sklearn. The default value is used for hyper-parameters not listed. }
\centering
\begin{tabular}{cc}
\toprule
\textbf{Hyper-parameter} & \textbf{Values} \\
\midrule
n\_estimators             & \{1000\} \\
max\_depth                & \{4, 6, 8, 10\} \\
min\_samples\_split        & \{5, 10, 15, 20\} \\
min\_samples\_leaf         & \{1, 3, 5, 8\} \\
random\_state             & \{42\} \\ 
bootstrap                & yes \\
max\_features             & sqrt \\
\bottomrule
\end{tabular}
\label{tab:rfparams}
\end{table}

\begin{table}[!ht]
 \caption{Example of expanding walk forward validation without where $t_{i}$ represents the feature vector of trading day $i$. In this example, a training window with an initial size $s=3$ is taken together with a testing window of size $k=1$. We therefore consistently use the feature vectors of past trading days to train a model (underlined) and subsequently test said model on trading day $t+k$ (bold). } 
  \centering
  \begin{tabular}{cc}
    \toprule
    Iteration & Variable roles\\
    \midrule
     1 & $ \underline{t_{1} \hspace{1.5mm} t_{2} \hspace{1.5mm} t_{3}} \hspace{1.5mm} \mathbf{ t_{4} } \hspace{1.5mm} t_{5} \hspace{1.5mm} \dotsi \hspace{1.5mm} t_{n} $ \\
     2 & $ \underline{ t_{1} \hspace{1.5mm} t_{2} \hspace{1.5mm} t_{3} \hspace{1.5mm} t_{4}} \hspace{1.5mm} \mathbf{t_{5}}  \hspace{1.5mm} \dotsi \hspace{1.5mm} t_{n} $ \\
     \vdots & \vdots \\
     m & $ \underline{ t_{1} \hspace{1.5mm} \dotsi \hspace{1.5mm} t_{n-3} \hspace{1.5mm} t_{n-2} \hspace{1.5mm} t_{n-1}} \hspace{1.5mm} \mathbf{t_{n}} $ \\

    \bottomrule
  \end{tabular}
  \label{ch3:tab:ev}
\end{table}

\subsection{Experimental Evaluation}
\label{meth:eval}
We evaluated the different random forest configurations using walk-forward validation, a cross-validation technique designed specifically for temporally ordered data. Classical cross-validation methods assume observations to be independent. This assumption does not necessarily hold for timeseries data, which inherently contains temporal dependencies among observations. To this end, the dataset is repeatedly split up in training and test sets where temporal order is accounted for. In our case, we used an expanding window of initially 504 trading days to train the models, after which performance was measured on the next out-of-sample 40 trading days. Table \ref{tab:ev} shows an example of this method where $t_{i}$ denotes the feature vector corresponding to trading day $i$. Note that in this scenario, when given a total of $n$ observations, an expanding training window of length $t$ and an out-of-sample test window of length $k$, you can construct a maximum of $n-t-k$ different train-test splits. Ultimately, each configuration its performance is averaged across all folds. We measured performance with the area under the receiver operating characteristic curve metric (AUC hereafter). 

\begin{table}[!ht]
 \caption{Example of walk-forward validation where $t_{i}$ represents the feature vector of trading day $i$. In this example, a training window with an initial size $s=3$ is taken together with a testing window of size $k=1$. We therefore consistently use the feature vectors of past trading days to train a model (underlined) and subsequently test said model on trading day $t+k$ (bold). } 
  \centering
  \begin{tabular}{cc}
    \toprule
    Iteration & Variable roles\\
    \midrule
     1 & $ \underline{t_{1} \hspace{1.5mm} t_{2} \hspace{1.5mm} t_{3}} \hspace{1.5mm} \mathbf{ t_{4} } \hspace{1.5mm} t_{5} \hspace{1.5mm} \dotsi \hspace{1.5mm} t_{n} $ \\
     2 & $ \underline{ t_{1} \hspace{1.5mm} t_{2} \hspace{1.5mm} t_{3} \hspace{1.5mm} t_{4}} \hspace{1.5mm} \mathbf{t_{5}}  \hspace{1.5mm} \dotsi \hspace{1.5mm} t_{n} $ \\
     \vdots & \vdots \\
     m & $ \underline{ t_{1} \hspace{1.5mm} \dotsi \hspace{1.5mm} t_{n-3} \hspace{1.5mm} t_{n-2} \hspace{1.5mm} t_{n-1}} \hspace{1.5mm} \mathbf{t_{n}} $ \\

    \bottomrule
  \end{tabular}
  \label{tab:ev}
\end{table}

\subsection{Analyzing Performance through Time with Hidden Markov Models}
\label{meth:hmm}
The ever-changing nature of financial markets makes it hard to find approaches that consistently work well. The ability to time approaches therefore becomes an interesting perk. In an effort to evaluate how our proposed approach weathers the evolution of the financial market through time, we look at its performance under different market regimes.

We quantified market regimes as different states in mean implied volatility using a hidden Markov model. For each stock, a different HMM model was trained on its end-of-day implied volatility timeseries dating from January 1st, 2007 till December 31st, 2012 and was used out-of-sample thereafter. Analogous to a study performed by Société Générale \citep{Socgen1}, four different regimes were identified corresponding to low, medium, high, and very high mean implied volatility. Table \ref{tab:hmmparams} specifies the hyper-parameter configuration used to build hidden Markov models using the hmmlearn Python package. 

\begin{table}[!htbp]
\setlength{\tabcolsep}{10pt} 
\caption{This table presents the hyper-parameter configuration used for building hidden Markov models using the hmmlearn package. The default value is used for hyper-parameters not listed. }
  \centering
\begin{tabular}{cc}
\toprule
\textbf{Hyper-parameter} & \textbf{Values} \\
\midrule
n\_components             & 4 \\
n\_iter                & 100 \\
random\_state          & 42 \\
emissions              & Gaussian \\
algorithm              & Viterbi \\
\bottomrule
\end{tabular}
\label{tab:hmmparams}
\end{table}

\section{Experimental Results and Discussion}
In this section, we present and discuss our experimental results. First, Section \ref{ch3:sec:abstud} shows the results of our ablation study where in total seven different feature configurations were considered. Section \ref{ch3:sec:sect} then builds on these results by looking at the performance of the best feature configuration per market sector. Lastly, Section \ref{ch3:sec:regim} looks at predictive performance across different implied volatility regimes. Recall that throughout the remainder of this section we may denote the 11 different stock market sectors by the symbol of their equivalent SPDR ETF tracker. This is solely done out of convenience and is not indicative of stocks only belonging to said ETF portfolio. Refer to Table \ref{tab:symb} for an overview of the sector symbols.

\subsection{Ablation Study}
\label{ch3:sec:abstud}
The first objective of this study was to investigate to what extent daily movements in end-of-day implied volatility can be predicted. To this end, we obtained 11 different features from three different data sources (Section \ref{meth:feats}) on which we built random forest classifiers to predict said target variable. We assessed the effectiveness of the different data sources by performing an ablation study where 7 different scenarios were considered, shown in Table \ref{tab:ablation}. In total, one-day ahead movements in implied volatility were predicted for 165 stocks (Section \ref{universe}) spanning an out-of-sample period from January 1st, 2013 till March 1st, 2019.  

\begin{table}[!ht]
 \caption{This tables shows the different feature scenarios considered in our ablation study together with their total number of features. Note that the third column indicates the usage of both original and derived features from the given feature source. } 
  \setlength{\tabcolsep}{10pt} 
  \centering
  \begin{tabular}{clc}
    \toprule
    Scenario & Feature Source & Features\\
    \midrule
    1 & Stock Price & 2 \\
    2 & Stock Price, Tweets & 8 \\
    3 & Implied Volatility & 3 \\
    4 & Implied Volatility, Tweets & 9 \\
    5 & Tweets & 6\\
    6 & Stock Price, Implied Volatility & 5 \\
    7 & Stock Price, Implied Volatility, Tweets & 11 \\
    \bottomrule
  \end{tabular}
  \label{tab:ablation}
\end{table}

We compared the predictive performance of different scenarios to that of a stratified dummy classifier, which makes the comparison more rigid than using a simple random classifier. Indeed, implied volatility tends to go down more often than it goes up. This causes a stratified dummy classifier to achieve a median AUC of 51.8\% across all 165 stocks versus a 50.0\% achieved by a fully random one. 

Table \ref{tab:abres} displays the median AUC achieved for each scenario averaged over the entire selected stock universe and the difference in AUC between our approach and the stratified dummy classifier. These results provide empirical evidence that end-of-day movements in implied volatility can indeed be predicted. All possible feature scenarios perform better than a purely random classifier that achieves a median of 50.0\% AUC. Moreover, 4 out of 7 scenarios outperform the stratified dummy classifier that achieves a median of 51.8\% AUC. The commonality among these improved scenarios is the use of implied volatility features, indicating that this is an important source of information. Moreover, including features derived from tweets always yielded a better median performance (S2 versus S1, S4 versus S3, and S7 versus S6). This implies there is indeed a predictive interplay between information on Twitter and future implied volatility. Lastly, using all possible features (S7) yielded the best result overall, suggesting there are predictive patterns among all three feature sources.

\begin{table}[!ht]
\setlength{\tabcolsep}{10pt} 
\caption{This table displays the median predictive performance across all 165 stocks per feature configuration obtained by predicting daily end-of-day movements in implied volatility over the period of January 1st, 2013 to March 1st, 2019. Moreover, the second row shows how much the proposed approach does better than the stratified dummy classifier. }
\centering
\begin{tabular}{cccccccc}
\toprule
                                  & S1 & S2 & S3 & S4 & S5 & S6 & S7 \\
\midrule
\multicolumn{1}{c}{Median AUC} & 51.1       & 51.6       & 53.6       & 54.2       & 50.9       & 54.3 & 55.1       \\
\multicolumn{1}{c}{Improvement} & -0.6          & \underline{-0.1}         & +1.9          & \underline{+2.5}          & -0.8         & +2.7 & \underline{\textbf{+3.4}} \\
\bottomrule
\end{tabular}
\label{tab:abres}
\end{table}

\subsection{Predictive Performance across Sectors}
\label{ch3:sec:sect}
\begin{figure}[ht!]
\centering
\includegraphics[scale=.35]{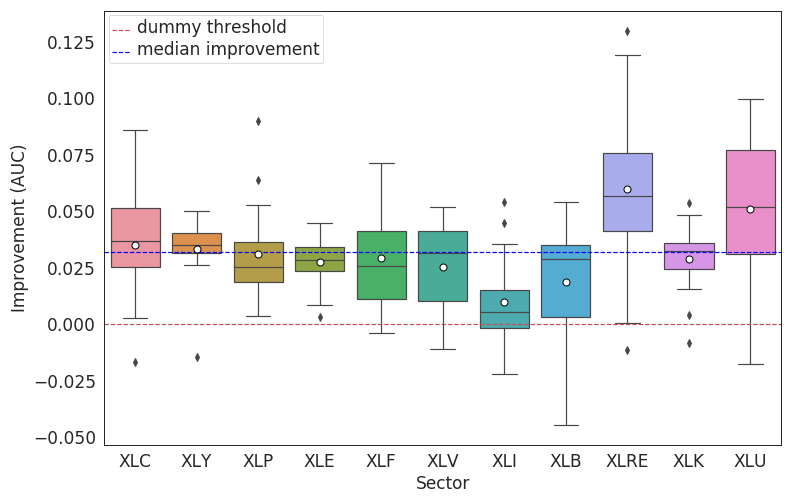}
\caption{This box plot shows the performance improvement of our proposed methodology using feature configuration S7 versus a stratified dummy classifier on individual stocks grouped by sector. The red dotted line represents the minimum threshold necessary to beat the stratified dummy classifier. The blue dotted line represents the overall median improvement of our proposed approach versus the stratified dummy classifier.}
\label{fig:secperf}
\end{figure}

In this section we look at the best performing feature configuration (S7) and its performance variability across 11 different stock market sectors. Figure \ref{fig:secperf} shows a box plot where the performance improvement of our proposed approach versus the stratified dummy classifier for each individual stock is grouped by sector. The results were obtained on an out-of-sample period spanning January 1st, 2013 till March 1st, 2019.

It is clear that our proposed methodology is generally able to beat the stratified dummy classifier across all different sectors. The approach beats the dummy classifier on 148 out of 165 stocks. However, there is a considerable amount of variability present across different sectors. Indeed, the approach does significantly better on XLRE and XLU, but predictions on XLC, XLY, and XLK also do better comparatively. In contrast, XLI and XLB seem to lack in performance. The next two subsections will showcase a preliminary attempt to partially explain this variability in performance.

\subsubsection{The Effect of Option Liquidity}
\begin{figure}[ht!]
\centering
\includegraphics[scale=.35]{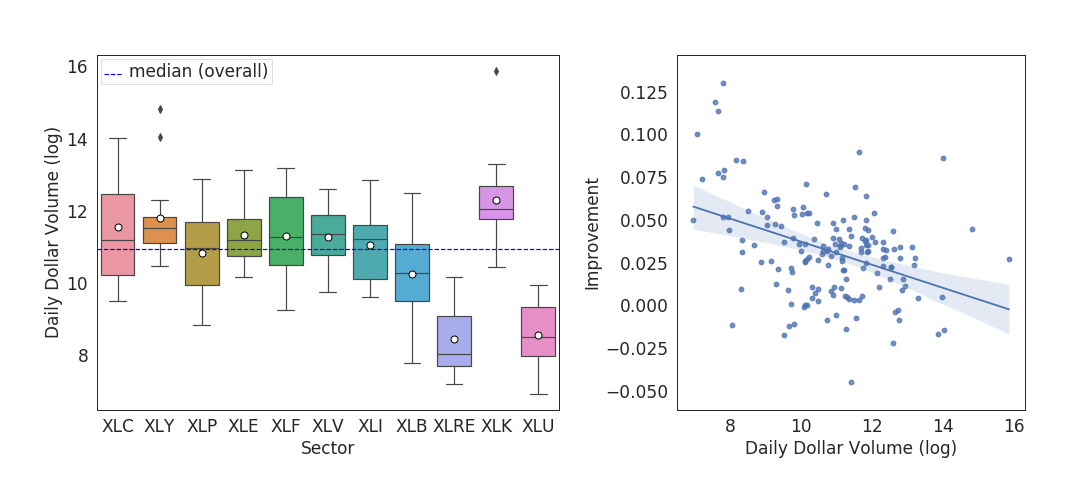}
\caption{This figure presents a box plot (left) where the median of daily stock option liquidity is grouped per sector and a regression plot (right) where the relationship between stock option liquidity and performance improvement versus the dummy classifier is outlined.}
\label{fig:secliq}
\end{figure}

The results from the previous section indicate that predictions on stocks from both XLRE and XLU do significantly better. Remarkably, it turns out that stocks in these two sectors are also significantly less liquid compared to other sectors. Figure \ref{fig:secliq} respectively shows a box plot of median option liquidity per sector, measured by the average daily dollar amount traded in options, and a regression plot where the relationship between liquidity and performance improvement versus the dummy classifier is outlined. 

These results seem to suggest that there is indeed a weak negative correlation between predictive performance and option liquidity, implying that less liquid stocks are easier to predict. 
We hypothesize that the relatively undersized liquidity of both XLRE and XLU in the options market is possibly accompanied by a less efficient price discovery process. This in turn might cause these markets to reflect new information more slowly, making them easier to predict with the information at hand. However, we note that is only one possible explanation and many other factors might lie at the basis of this phenomenon. 

\subsubsection{The Effect of Twitter Attention}
\begin{figure}[ht!]
\centering
\includegraphics[scale=.35]{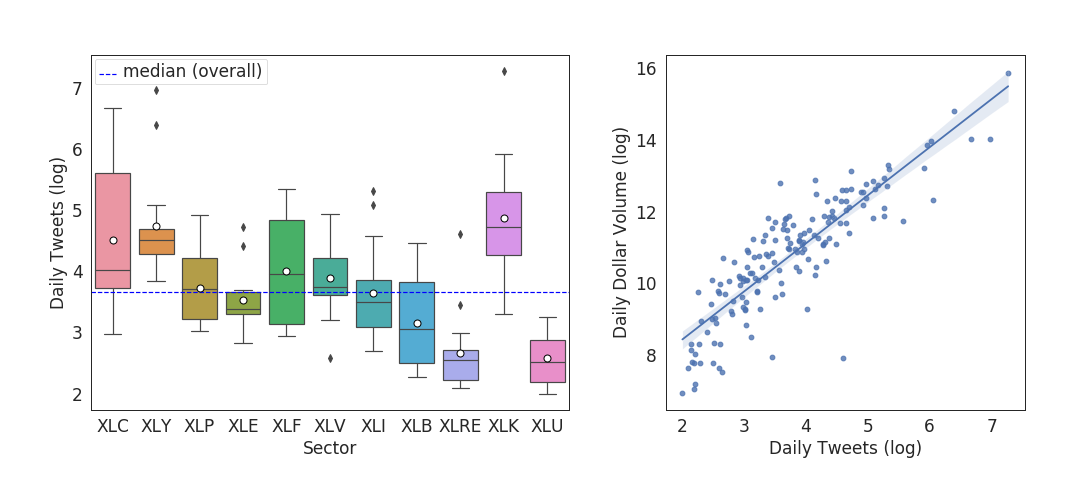}
\caption{This figure presents a box plot (left) where the median of daily tweet publication of stocks is grouped per sector and a regression plot (right) where the relationship between a stock its daily tweets and option liquidity is outlined.}
\label{fig:tweetliq}
\end{figure}

Lower liquidity might partially explain why sectors such as XLRE and XLU seem easier to predict, but it certainly does not tell the whole story. Indeed, predictions on stocks from XLC, XLY, and XLK, examples of very liquid sectors, also do better comparatively. Here, we hypothesize that this might be due to the attention they receive on Twitter. Figure \ref{fig:tweetliq} respectively shows a box plot of the median daily tweets published on stocks grouped per sector and a regression plot where the relationship between liquidity and daily tweets is outlined. 

\begin{figure}[ht!]
\centering
\includegraphics[scale=.35]{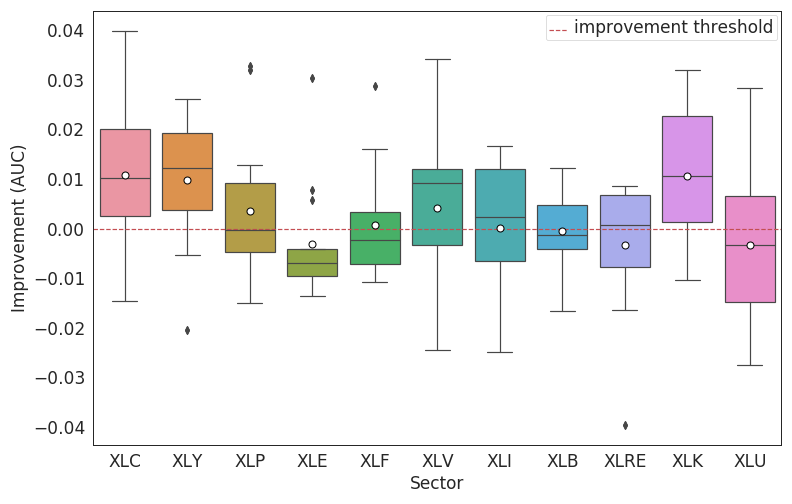}
\caption{This figure shows the performance improvement of using Twitter features together with features from S6, versus only using features from S6. The red dotted line represents the minimum threshold necessary to beat the feature configuration of S6.}
\label{fig:tweetimprov}
\end{figure}

The box plot on the left of Figure \ref{fig:tweetliq} shows that stocks in XLC, XLY, and XLK receive significantly more attention on Twitter than others. Note that the plot demonstrates a striking resemblance with the box plot showing liquidity per sector in Figure \ref{fig:secliq}. Indeed, the regression plot on the right shows a very strong correlation between attention on Twitter and liquidity. These findings seem to suggest that prediction is easier on stocks that are more popular on social media. We investigated this phenomenon further by looking at the improvement in predictive performance caused by features extracted from Twitter per sector. More concrete, we looked at the difference in performance between feature configuration S6, which uses stock and options features, and S7, which combines features from S6 and Twitter features (Section \ref{ch3:sec:abstud}). Figure \ref{fig:tweetimprov} shows the median improvement of using Twitter features per stock grouped by sector.  

With exception of XLE, most sectors seem to have a sizeable number of stocks that benefit from using social media features. This is no surprise considering the results presented in Section \ref{ch3:sec:abstud}. However, it seems that sectors that receive more social media attention seem to benefit the most. For example, the three sectors XLC, XLY, and XLK that are most popular also benefit more consistently followed closely by XLV. Two reasons might explain these results. First, previous research has hinted at social media inciting herd behavior and emotional reactions among investors, possibly driving inefficiency \citep[e.g.][]{herding, Oliveira2017, GROKLUMANN2019171}. If this is true, it makes sense that Twitter features provide more added value for popular stocks. Second, our sentiment extraction technique on tweets is imperfect and might impact these results as well. Without carefully filtering out tweets that are advertisements or spam, we rely on the law of large numbers to correctly estimate a stock its sentiment for a given day. Naturally, daily sentiment estimation will therefore be better on stocks that are more heavily tweeted about. 

\subsection {Performance across Implied Volatility Regimes}
\label{ch3:sec:regim}

In this section, we look at the best-performing feature configuration (S7) and its performance variability across four different market regimes in implied volatility, identified by using a hidden Markov model. Table \ref{tab:reg1} shows the median of all results across 165 stocks where, for each of the four regimes, the columns respectively show how many days were spent in a regime, what the average implied volatility was, the AUC of a stratified dummy classifier, and how much better our approach did compared to the latter. 


\begin{table}[!ht]
\centering
\setlength{\tabcolsep}{6pt}
\caption{This table shows the median number of days a stock resided in one out of four implied volatility regimes, together with each regime's mean implied volatility, stratified dummy performance and the improvement of our approach. In total, 165 stocks were considered over a period spanning January 1st, 2013 till March 1st, 2019. }
\begin{tabular}{cccccc}
\toprule
          & Frequency & Implied Volatility & Dummy AUC &  Improvement \\
\midrule
Low       & 392       & 18.6               & 50.75     & +3.15 &      \\
Medium    & 452       & 22.3               & 50.65     & +3.83 &      \\
High      & 411       & 26.7               & 52.12     & +3.84 &      \\
Very High & 231       & 35.3               & 54.92     & +1.47 &      \\
\midrule
\end{tabular}
\label{tab:reg1}
\end{table}

In general, stocks seem to reside more in the lower implied volatility environments. The stratified dummy classifier also appears to perform worse here, implying that up and down movements are about equal in occurrence. However, the dummy classifier performance picks up as implied volatility increases. This makes sense from a theoretical perspective, as implied volatility is deemed to be a mean reverting process characterized by big up movements after which the variable slowly trails back down, resulting in more down movements. Although our approach seems to outperform the dummy classifier in all regimes, the added value seems to be most significant in the low to high regimes. This may suggest that distressed markets are harder to predict than their calmer counterparts. However, note that the improvement of our approach seems to be slightly correlated with regime frequency. From a data science perspective, this might imply that the model performs worse in less frequent regimes because it simply had fewer examples to learn from. Lastly, Table \ref{tab:reg2} offers a more finer-grained analysis where the median improvement for each regime per sector is shown.

\begin{table}[!ht]
\centering
\setlength{\tabcolsep}{6pt}
\caption{This table shows the median improvement of our approach compared to a stratified dummy classifier for stocks grouped per market sector and implied volatility regime. The best and worst regime for each sector are respectively indicated by bold and underlined text styles.}
\begin{tabular}{ccccc}
\toprule
          & Low & Medium & High &  Very High \\
\midrule
XLB & 1.87 &	\textbf{3.06} &	1.02 &	\underline{0.77} \\
XLC & 3.57 &	4.76 &	\textbf{5.27} &	\underline{0.26} \\
XLE & 2.55 &	2.46 &	\textbf{4.0} &	\underline{-2.55} \\
XLF & 3.06 &	\textbf{4.08} &	2.64 &	\underline{-1.7} \\
XLI & 0.05 &	\textbf{1.53} &	0.17 &	\underline{-1.5} \\
XLK & \underline{0.60} &	2.38 &	3.91 &	\textbf{5.70} \\
XLP & 2.72 &	\textbf{3.83} &	3.23 &	\underline{1.53} \\
XLRE & \textbf{8.5} &	6.38 &	4.93 &	\underline{4.25} \\
XLU & \textbf{6.63} &	5.95 &	5.36 &	\underline{1.79} \\
XLV & \textbf{4.25} &	3.23 &	\underline{1.62} &	2.13 \\
XLY & 2.04 &	3.66 &	\textbf{5.95} & \underline{0.85} \\
\midrule
\end{tabular}
\label{tab:reg2}
\end{table}

Again, we remark a significant amount of variability in performance across both regimes and sectors. Optimal implied volatility regimes seem to differ significantly for different sectors. In contrast, with exception of XLK, sectors seem to comparatively do worse when implied volatility is very high. As all sectors share roughly the same regime frequency distribution, no clear reason emerges to explain the performance variability. One potential reason might be due to sector idiosyncrasies. For example, defensive sectors such as XLRE, XLU, and XLV seem to have lower optimal regimes than cyclical sectors such as XLC, XLK, and XLY. However, this is not always the case. Lastly, we have no explanation for the significant difference between XLK and the other sectors. Here, our approach performs best in the highest implied volatility regime and worst in the lowest. This is in stark contrast with the other sectors where the opposite is true. Perhaps the speculative nature of technology stocks is more sensitive to herd behaviour and therefore investor irrationality.

\section{Conclusion}
In this study, we presented the first empirical evidence that one-day ahead movements of end-of-day stock implied volatility can be predicted to a certain extent, and that attention and sentiment features extracted from Twitter improve the performance of the approach. These alternative features were not able to predict implied volatility in isolation, but improved the approach significantly when combined with predictors extracted from stock and options data. This suggests that the interplay between these sources gives rise to predictive patterns. By conducting our experiments on a diversified universe of 165 US stocks, we were able to assess the predictive performance across 11 traditional stock market sectors and found that stocks in real estate, utilities, consumer discretionary, communications, and technology were easier to predict than others. Further analysis indicated that these differences could potentially be explained by market inefficiencies caused by low option liquidity in the real estate and utilities sector, and excessive Twitter attention in the consumer discretionary, communications, and technology sector. Lastly, using hidden Markov models, we evaluated the predictive performance of our approach across four different implied volatility regimes. Although it outperforms the dummy classifier in all four regimes, we found that it yields the least improvement in the regime associated with the highest average implied volatility. Moreover, we discovered that different stock market sectors have different optimal regimes for the application of our approach. By analyzing performance through the usage of regimes, we showed that this alternative quantification of time provides additional insight into the performance of models which in turn could help better anticipate their future performance.

\appendix

\clearpage
\section{Selected Stock Universe}
\label{appA}

This appendix contains more detailed information on the US stock universe used in this study. Table \ref{tab:stockuniv} outlines which stocks were chosen for each traditional market sector.

\begin{table}[ht!]
\centering
\setlength{\tabcolsep}{6pt}
\caption{This table presents the US stock universe that was used in this study. Note that for each traditional market sector we chose 15 stocks based on option liquidity.}
\begin{tabular}{cccccc}
\toprule
\multirow{3}*{Materials} & FCX & X & NEM & CLF & MOS \\
                        & DD & IP & CF & AA & NUE \\
                        & LYB & VMC & SHW & BLL & WRK \\
\midrule
\multirow{3}*{Communications} & FB & NFLX & GOOG & T & TWTR \\
                        & VZ & DIS & CMCSA & EA & YELP \\
                        & ATVI & DISH & CHTR & Z & TTWO \\
\midrule
\multirow{3}*{Energy} & XOM & OXY & CVX & COP & SLB \\
                        & HAL & VLO & EOG & APA & KMI \\
                        & HES & MPC & MRO & RIG & WMB \\
\midrule
\multirow{3}*{Financials} & BAC & JPM & C & GS & WFC \\
                        & AIG & BX & MS & AXP & CME \\
                        & MET & USB & SCHW & COF & BLK \\
\midrule
\multirow{3}*{Industrials} & GE & BA & CAT & LMT & UPS \\
                        & UNP & FDX & DE & MMM & HON \\
                        & CSX & NSC & EMR & NOC & ETN \\
\midrule
\multirow{3}*{Technology} & AAPL & NVDA & MSFT & INTC & MU \\
                        & IBM & QCOM & CSCO & CRM & AMD \\
                        & MA & V & ORCL & ADBE & TXN \\
\midrule
\multirow{3}*{C. Staples} & PG & WMT & PM & KO & MO \\
                        & CL & COST & PEP & HLF & WBA \\
                        & GIS & MNST & TSN & CAG & KR \\
\midrule
\multirow{3}*{Real Estate} & SPG & WY & AMT & EQIX & IRM \\
                        & CCI & PSA & AVB & VTR & O \\
                        & HST & DLR & PLD & MAC & EQR \\
\midrule
\multirow{3}*{Utilities} & SO & EXC & AEP & DUK & NRG \\
                        & NEE & FE & D & PPL & ED \\
                        & ETR & EIX & CNP & NI & SRE \\
\midrule
\multirow{3}*{Healthcare} & PFE & JNJ & GILD & LLY & MRK \\
                        & ABT & BMY & AMGN & UNH & CVS \\
                        & ABBV & ISRG & MDT & CI & DHR \\
\midrule
\multirow{3}*{C. Discretionary} & AMZN & TSLA & MCD & HD & F \\
                        & CMG & GM & SBUX & EBAY & NKE \\
                        & TGT & LOW & BBY & LULU & MGM \\
\bottomrule
\end{tabular}
\label{tab:stockuniv}
\end{table}

\bibliographystyle{elsarticle-harv} 
\bibliography{main}

\begin{thebibliography}{24}
\expandafter\ifx\csname natexlab\endcsname\relax\def\natexlab#1{#1}\fi
\providecommand{\url}[1]{\texttt{#1}}
\providecommand{\href}[2]{#2}
\providecommand{\path}[1]{#1}
\providecommand{\DOIprefix}{doi:}
\providecommand{\ArXivprefix}{arXiv:}
\providecommand{\URLprefix}{URL: }
\providecommand{\Pubmedprefix}{pmid:}
\providecommand{\doi}[1]{\href{http://dx.doi.org/#1}{\path{#1}}}
\providecommand{\Pubmed}[1]{\href{pmid:#1}{\path{#1}}}
\providecommand{\bibinfo}[2]{#2}
\ifx\xfnm\relax \def\xfnm[#1]{\unskip,\space#1}\fi
\bibitem[{Baker and Wurgler(2006)}]{baker2006}
\bibinfo{author}{Baker, M.}, \bibinfo{author}{Wurgler, J.},
  \bibinfo{year}{2006}.
\newblock \bibinfo{title}{Investor sentiment and the cross-section of stock
  returns}.
\newblock \bibinfo{journal}{The Journal of Finance} \bibinfo{volume}{61},
  \bibinfo{pages}{1645--1680}.
\newblock \DOIprefix\doi{https://doi.org/10.1111/j.1540-6261.2006.00885.x}.
\bibitem[{Baum(1972)}]{baumwelch}
\bibinfo{author}{Baum, L.E.}, \bibinfo{year}{1972}.
\newblock \bibinfo{title}{An inequality and associated maximization technique
  in statistical estimation for probabilistic functions of {M}arkov processes},
  in: \bibinfo{booktitle}{Inequalities {III}: {P}roceedings of the Third
  Symposium on Inequalities}, \bibinfo{publisher}{Academic Press}. pp.
  \bibinfo{pages}{1--8}.
\bibitem[{Bollen et~al.(2011)Bollen, Mao and Zeng}]{bollen}
\bibinfo{author}{Bollen, J.}, \bibinfo{author}{Mao, H.}, \bibinfo{author}{Zeng,
  X.}, \bibinfo{year}{2011}.
\newblock \bibinfo{title}{The impact of social and conventional media on firm
  equity value: A sentiment analysis approach}.
\newblock \bibinfo{journal}{Journal of Computational Science}
  \bibinfo{volume}{2}, \bibinfo{pages}{1--8}.
\newblock \DOIprefix\doi{https://doi.org/10.1016/j.jocs.2010.12.007}.
\bibitem[{Breiman(2001)}]{rf}
\bibinfo{author}{Breiman, L.}, \bibinfo{year}{2001}.
\newblock \bibinfo{title}{Random forests}.
\newblock \bibinfo{journal}{Mach. Learn.} \bibinfo{volume}{45},
  \bibinfo{pages}{5–32}.
\newblock \DOIprefix\doi{https://doi.org/10.1023/A:1010933404324}.
\bibitem[{Breiman et~al.(1984)Breiman, Friedman, Olshen and Stone}]{cart}
\bibinfo{author}{Breiman, L.}, \bibinfo{author}{Friedman, J.H.},
  \bibinfo{author}{Olshen, R.A.}, \bibinfo{author}{Stone, C.J.},
  \bibinfo{year}{1984}.
\newblock \bibinfo{title}{Classification and Regression Trees}.
\newblock \bibinfo{publisher}{Wadsworth and Brooks},
  \bibinfo{address}{Monterey, CA}.
\newblock \DOIprefix\doi{https://doi.org/10.1201/9781315139470}.
\bibitem[{Caruana et~al.(2008)Caruana, Karampatziakis and
  Yessenalina}]{boostbest}
\bibinfo{author}{Caruana, R.}, \bibinfo{author}{Karampatziakis, N.},
  \bibinfo{author}{Yessenalina, A.}, \bibinfo{year}{2008}.
\newblock \bibinfo{title}{An empirical evaluation of supervised learning in
  high dimensions}, in: \bibinfo{booktitle}{International Conference on Machine
  Learning (ICML)}, pp. \bibinfo{pages}{96--103}.
\newblock \DOIprefix\doi{https://doi.org/10.1145/1390156.1390169}.
\bibitem[{Caruana and Niculescu-Mizil(2006)}]{boostbest2}
\bibinfo{author}{Caruana, R.}, \bibinfo{author}{Niculescu-Mizil, A.},
  \bibinfo{year}{2006}.
\newblock \bibinfo{title}{An empirical comparison of supervised learning
  algorithms}, in: \bibinfo{booktitle}{Proceedings of the 23rd International
  Conference on Machine Learning}, \bibinfo{publisher}{Association for
  Computing Machinery}, \bibinfo{address}{New York, NY, USA}. p.
  \bibinfo{pages}{161–168}.
\newblock \DOIprefix\doi{https://doi.org/10.1145/1143844.1143865}.
\bibitem[{CBOE(2015)}]{VIX}
\bibinfo{author}{CBOE}, \bibinfo{year}{2015}.
\newblock \bibinfo{title}{Cboe volatility index white paper} \URLprefix
  \url{https://www.cboe.com/micro/vix/vixwhite.pdf}.
\bibitem[{{Chang} and {Hancock}(1966)}]{fwdback}
\bibinfo{author}{{Chang}, R.}, \bibinfo{author}{{Hancock}, J.},
  \bibinfo{year}{1966}.
\newblock \bibinfo{title}{On receiver structures for channels having memory}.
\newblock \bibinfo{journal}{IEEE Transactions on Information Theory}
  \bibinfo{volume}{12}, \bibinfo{pages}{463--468}.
\newblock \DOIprefix\doi{https://doi.org/10.1109/TIT.1966.1053923}.
\bibitem[{Chiang and Zheng(2010)}]{herd2}
\bibinfo{author}{Chiang, T.C.}, \bibinfo{author}{Zheng, D.},
  \bibinfo{year}{2010}.
\newblock \bibinfo{title}{An empirical analysis of herd behavior in global
  stock markets}.
\newblock \bibinfo{journal}{Journal of Banking \& Finance}
  \bibinfo{volume}{34}, \bibinfo{pages}{1911--1921}.
\newblock \DOIprefix\doi{https://doi.org/10.1016/j.jbankfin.2009.12.014}.
\bibitem[{Daviaud et~al.(2020)Daviaud, Korber, Mukhopadhyay and
  Ungari}]{Socgen1}
\bibinfo{author}{Daviaud, O.}, \bibinfo{author}{Korber, O.},
  \bibinfo{author}{Mukhopadhyay, A.}, \bibinfo{author}{Ungari, S.},
  \bibinfo{year}{2020}.
\newblock \bibinfo{title}{Systematic trading in options}.
\newblock \bibinfo{journal}{Société Générale - Cross Asset Research} .
\bibitem[{Fama(1970)}]{fama}
\bibinfo{author}{Fama, E.F.}, \bibinfo{year}{1970}.
\newblock \bibinfo{title}{Efficient capital markets: A review of theory and
  empirical work}.
\newblock \bibinfo{journal}{The Journal of Finance} \bibinfo{volume}{25},
  \bibinfo{pages}{383--417}.
\newblock \DOIprefix\doi{https://doi.org/10.2307/2325486}.
\bibitem[{{Forney}(1973)}]{viterbi}
\bibinfo{author}{{Forney}, G.D.}, \bibinfo{year}{1973}.
\newblock \bibinfo{title}{The viterbi algorithm}.
\newblock \bibinfo{journal}{Proceedings of the IEEE} \bibinfo{volume}{61},
  \bibinfo{pages}{268--278}.
\newblock \DOIprefix\doi{https://doi.org/10.1109/PROC.1973.9030}.
\bibitem[{Friedman(2001)}]{gbm}
\bibinfo{author}{Friedman, J.H.}, \bibinfo{year}{2001}.
\newblock \bibinfo{title}{{Greedy function approximation: A gradient boosting
  machine.}}
\newblock \bibinfo{journal}{The Annals of Statistics} \bibinfo{volume}{29},
  \bibinfo{pages}{1189 -- 1232}.
\newblock \DOIprefix\doi{https://doi.org/10.1214/aos/1013203451}.
\bibitem[{Groß-Klußmann et~al.(2019)Groß-Klußmann, König and
  Ebner}]{GROKLUMANN2019171}
\bibinfo{author}{Groß-Klußmann, A.}, \bibinfo{author}{König, S.},
  \bibinfo{author}{Ebner, M.}, \bibinfo{year}{2019}.
\newblock \bibinfo{title}{Buzzwords build momentum: Global financial twitter
  sentiment and the aggregate stock market}.
\newblock \bibinfo{journal}{Expert Systems with Applications}
  \bibinfo{volume}{136}, \bibinfo{pages}{171--186}.
\newblock \DOIprefix\doi{https://doi.org/10.1016/j.eswa.2019.06.027}.
\bibitem[{Guijarro et~al.(2019)Guijarro, Moya-Clemente and Saleemi}]{twitliq}
\bibinfo{author}{Guijarro, F.}, \bibinfo{author}{Moya-Clemente, I.},
  \bibinfo{author}{Saleemi, J.}, \bibinfo{year}{2019}.
\newblock \bibinfo{title}{{Liquidity Risk and Investors’ Mood: Linking the
  Financial Market Liquidity to Sentiment Analysis through Twitter in the
  S\&P500 Index}}.
\newblock \bibinfo{journal}{Sustainability} \bibinfo{volume}{11},
  \bibinfo{pages}{1--13}.
\newblock \DOIprefix\doi{https://doi.org/10.3390/su11247048}.
\bibitem[{Hutto and Gilbert(2014)}]{vader}
\bibinfo{author}{Hutto, C.J.}, \bibinfo{author}{Gilbert, E.},
  \bibinfo{year}{2014}.
\newblock \bibinfo{title}{Vader: A parsimonious rule-based model for sentiment
  analysis of social media text.}, in: \bibinfo{editor}{Adar, E.},
  \bibinfo{editor}{Resnick, P.}, \bibinfo{editor}{Choudhury, M.D.},
  \bibinfo{editor}{Hogan, B.}, \bibinfo{editor}{Oh, A.H.} (Eds.),
  \bibinfo{booktitle}{ICWSM}, \bibinfo{publisher}{The AAAI Press}.
\bibitem[{Karagozoglu and Fabozzi(2017)}]{Karagozoglu}
\bibinfo{author}{Karagozoglu, A.K.}, \bibinfo{author}{Fabozzi, F.J.},
  \bibinfo{year}{2017}.
\newblock \bibinfo{title}{Volatility wisdom of social media crowds}.
\newblock \bibinfo{journal}{The Journal of Portfolio Management}
  \bibinfo{volume}{43}, \bibinfo{pages}{136--151}.
\newblock \DOIprefix\doi{https://doi.org/10.3905/jpm.2017.43.2.136}.
\bibitem[{{Khoshgoftaar} et~al.(2011){Khoshgoftaar}, {Van Hulse} and
  {Napolitano}}]{noisebag}
\bibinfo{author}{{Khoshgoftaar}, T.M.}, \bibinfo{author}{{Van Hulse}, J.},
  \bibinfo{author}{{Napolitano}, A.}, \bibinfo{year}{2011}.
\newblock \bibinfo{title}{Comparing boosting and bagging techniques with noisy
  and imbalanced data}.
\newblock \bibinfo{journal}{IEEE Transactions on Systems, Man, and Cybernetics
  - Part A: Systems and Humans} \bibinfo{volume}{41},
  \bibinfo{pages}{552--568}.
\newblock \DOIprefix\doi{https://doi.org/10.1109/TSMCA.2010.2084081}.
\bibitem[{Li et~al.(2018)Li, van Dalen and van Rees}]{twitnoise}
\bibinfo{author}{Li, T.}, \bibinfo{author}{van Dalen, J.}, \bibinfo{author}{van
  Rees, P.J.}, \bibinfo{year}{2018}.
\newblock \bibinfo{title}{More than just noise? examining the information
  content of stock microblogs on financial markets}.
\newblock \bibinfo{journal}{Journal of Information Technology}
  \bibinfo{volume}{33}, \bibinfo{pages}{50--69}.
\newblock \DOIprefix\doi{https://doi.org/10.1057/s41265-016-0034-2}.
\bibitem[{Oliveira et~al.(2017)Oliveira, Cortez and Areal}]{Oliveira2017}
\bibinfo{author}{Oliveira, N.}, \bibinfo{author}{Cortez, P.},
  \bibinfo{author}{Areal, N.}, \bibinfo{year}{2017}.
\newblock \bibinfo{title}{The impact of microblogging data for stock market
  prediction: using twitter to predict returns, volatility, trading volume and
  survey sentiment indices}.
\newblock \bibinfo{journal}{Expert systems with applications}
  \bibinfo{volume}{73}, \bibinfo{pages}{125--144}.
\newblock \DOIprefix\doi{https://doi.org/10.1016/j.eswa.2016.12.036}.
\bibitem[{{Rabiner} and {Juang}(1986)}]{hmm}
\bibinfo{author}{{Rabiner}, L.}, \bibinfo{author}{{Juang}, B.},
  \bibinfo{year}{1986}.
\newblock \bibinfo{title}{An introduction to hidden markov models}.
\newblock \bibinfo{journal}{IEEE ASSP Magazine} \bibinfo{volume}{3},
  \bibinfo{pages}{4--16}.
\newblock \DOIprefix\doi{https://doi.org/10.1109/MASSP.1986.1165342}.
\bibitem[{Schnaubelt et~al.(2020)Schnaubelt, Fischer and
  Krauss}]{SCHNAUBELT2020103895}
\bibinfo{author}{Schnaubelt, M.}, \bibinfo{author}{Fischer, T.G.},
  \bibinfo{author}{Krauss, C.}, \bibinfo{year}{2020}.
\newblock \bibinfo{title}{Separating the signal from the noise – financial
  machine learning for twitter}.
\newblock \bibinfo{journal}{Journal of Economic Dynamics and Control}
  \bibinfo{volume}{114}, \bibinfo{pages}{103895}.
\newblock \DOIprefix\doi{https://doi.org/10.1016/j.jedc.2020.103895}.
\bibitem[{Wang and Wang(2018)}]{herding}
\bibinfo{author}{Wang, G.}, \bibinfo{author}{Wang, Y.}, \bibinfo{year}{2018}.
\newblock \bibinfo{title}{Herding, social network and volatility}.
\newblock \bibinfo{journal}{Economic Modelling} \bibinfo{volume}{68},
  \bibinfo{pages}{74--81}.
\newblock \DOIprefix\doi{https://doi.org/10.1016/j.econmod.2017.04.018}.

\end{thebibliography}





\end{document}